\newif\ifSHOWPROOFS
\documentclass[a4paper]{llncs}
\SHOWPROOFSfalse
\SHOWPROOFStrue

\usepackage{a4,a4wide}
\usepackage{xspace}
\usepackage{times}
\usepackage{url}
\usepackage{amsmath,amssymb}
\usepackage{algorithm}
\usepackage{algorithmic}
\usepackage{graphicx}
\usepackage{pst-tree}
\usepackage{pst-node}
\usepackage{epstopdf}
\newcommand{\ruleimp}{\mathtt{{:-}}}
\renewcommand{\ruleimp}{:\!- \ }

\newcommand{\naf}{\ensuremath{\mathrm{not}}\xspace}

\newcommand{\OntoStudio}{\texttt{OntoStudio}\xspace} 
\newcommand{\OntoDLV}{\texttt{OntoDLV}\xspace}

\newcommand{\gringo}{\texttt{Gringo}\xspace}
\newcommand{\aspide}{\texttt{ASPIDE}\xspace}
\newcommand{\lparse}{\texttt{Lparse}\xspace}
\newcommand{\aspviz}{\texttt{ASPVIZ}\xspace}
\newcommand{\idpdraw}{\texttt{IDPDraw}\xspace}
\newcommand{\igrom}{\texttt{iGROM}\xspace}
\newcommand{\ape}{\texttt{APE}\xspace}
\newcommand{\clasp}{\texttt{Clasp}\xspace}
\newcommand{\dlv}{\texttt{DLV}\xspace}

\newcommand{\sealion}{\texttt{SeaLion}\xspace}
\newcommand{\kara}{\texttt{Kara}\xspace}

\newcommand{\abdtrans}[2]{\ensuremath{\lambda({#1},{#2})}\xspace}

\newcommand{\be}{\begin{compactenum}}
\newcommand{\ee}{\end{compactenum}}
\newcommand{\bi}{\begin{compactitem}}
\newcommand{\ei}{\end{compactitem}}

\newcommand{\iec}[0]{i.e.,\xspace}

\newcommand{\egc}[0]{e.g.,\xspace}

\newcommand{\nop}[1]{}

\title{The \sealion has Landed:
       An IDE for Answer-Set Programming---Preliminary Report%
\thanks{This work was partially supported by the Austrian Science Fund (FWF)
under project P21698.}}

\author{%
Johannes Oetsch \and J\"org P\"uhrer \and Hans Tompits}

\institute{%
Institut f\"ur Informationssysteme 184/3,\\
Technische Universit\"at Wien,\\
Favoritenstra\ss{}e\ 9-11,
A-1040 Vienna, Austria \\
\email{\{oetsch,puehrer,tompits\}@kr.tuwien.ac.at}
}

\begin{document}

\maketitle
                                              
\begin{abstract}
We report about the current state and designated features of the tool \sealion, aimed to serve as an integrated development environment (IDE) for
answer-set programming (ASP).
A main goal of \sealion is to provide
a user-friendly environment for supporting a developer to
write, evaluate, debug, and test  answer-set programs.
To this end, new support techniques have to be developed that suit
the requirements of the answer-set semantics and meet
the constraints of practical applicability.
In this respect, \sealion benefits from the research results of 
a project on methods and methodologies
for answer-set program development 
in whose context \sealion is realised.
Currently, the tool provides source-code editors for the languages of \gringo and \dlv
that offer syntax highlighting, syntax checking, and a visual program outline.
Further implemented features are support for external solvers and 
visualisation as well as visual editing of answer sets.
\sealion comes as a plugin of the popular Eclipse platform
and provides itself interfaces for future extensions of the IDE.
\end{abstract}

\section{Introduction}\label{sec:intro}
Answer-set programming (ASP) is a well-known and fully declarative problem-solving paradigm based on the idea that solutions to computational problems are represented in terms of logic programs such that the models of the latter, referred to as the \emph{answer sets}, 
provide the solutions of a problem instance.\footnote{For an overview about ASP, we refer the reader to a survey article by Gelfond and Leone~\cite{GelfondL02} or the textbook by Baral~\cite{baral03}.}
In recent years, the expressibility of languages supported by answer-set solvers
increased significantly~\cite{clasp09}.
As well, ASP solvers have become much more efficient,
\egc the solver \clasp proved to be competitive with state-of-the-art SAT solvers~\cite{satcomp09}.

Despite these improvements in solver technology,
a lack of suitable \emph{engineering tools} for {developing} programs
is still a handicap for ASP towards gaining widespread popularity as a problem-solving paradigm.
This issue is clearly recognised in the ASP community and work to fill this gap has started recently, addressing issues like debugging, testing, and the modularity of programs~\cite{BrainV05,pontelli09,debinc,brai-etal-07,WittocxVD09,Oetsch-etal09-debug,ecai10,testeval11,JOTW09:jair}.
Additionally, in order to facilitate tool support as known for other programming languages, attempts to provide \emph{integrated development environments} (IDEs) have been put forth. 
Work in this direction includes the systems \ape~\cite{apesea07}, \aspide~\cite{aspide11}, and \igrom~\cite{igrom}.

Following this endeavour, in this paper, we describe the current status and designated features of a further IDE, \sealion, developed as part of an ongoing research project on methods and methodologies for developing answer-set programs~\cite{mmdasp}.

\sealion is designed as an Eclipse plugin, providing useful and intuitive features for ASP.
Besides experts,
the target audience for \sealion are software developers new to ASP, yet who are familiar with support tools as used in procedural and object-oriented programming.
Our goal is to fully support the languages of the current state-of-the-art solvers
\clasp (in conjunction with \gringo)~\cite{clasp09,gringo07} and \dlv~\cite{leone06}, which distinguishes \sealion from the other IDEs mentioned above which support only a single solver. Indeed,
\ape~\cite{apesea07}, which is also an Eclipse plugin, supports only the language of \lparse~\cite{lparse} that is a subset
of the language of \gringo, whilst \aspide~\cite{aspide11}, a recently developed standalone IDE, offers support only for \dlv programs.
Although \igrom provides basic functionality for the languages of both \lparse and \dlv~\cite{igrom}, it currently does not support the latest version of \dlv or the full syntax of \gringo.

At present, \sealion is in an alpha version that already
  implements important core functionality.
In particular, the languages of \dlv and \gringo are supported to a large extent.
The individual parsers translate programs and answer sets to
data structures that are part of a rich and flexible framework for 
internally representing program elements.
Based on these structures,
the editor provides syntax highlighting,
syntax checks, error reporting, error highlighting,
and automatic generation of a program outline.
There is functionality to manage external tools such as answer-set solvers
and to define arbitrary pipes between them (as needed when using separate grounders and solvers).
Moreover, in order to run an answer-set solver on the created programs, launch configurations can be created
in which the user can choose input files, a solver configuration,  command line arguments for the solver, as well as output-processing strategies.
Answer sets resulting from a launch can either be parsed and stored in a view for interpretations,
or the solver output can be displayed unmodified in Eclipse's built-in console view.

Another key feature of \sealion is the capability for the
\emph{visualisation} and \emph{visual editing} of interpretations.
This follows ideas from the  visualisation tools
\aspviz~\cite{CliffeVBP08} and \idpdraw~\cite{idpdraw}, where a visualisation program $\Pi_V$ (itself being an answer-set program) is joined with an interpretation $I$ 
that shall be visualised.
Subsequently, the overall program is evaluated using an answer-set solver, and the visualisation is generated from a resulting answer set. 
However, the editing feature of \sealion allows also to graphically manipulate the interpretations under consideration which is not supported by \aspviz and \idpdraw.

The visualisation functionality of \sealion is itself represented as an Eclipse plugin, called \kara.\footnote{The name derives, with all due respect, from ``Kara Zor-El'', the native Kryptonian name of \emph{Supergirl}, given that Kryptonians have visual superpowers on Earth.}
In this paper, however, we describe only the basic functionality of \kara; a full description  is given in a companion paper~\cite{kara}.

%%%%%%%%%%%%%%%%%%%%%%%%%%%%%%%%%%%%%%%%%%%%%%%%%%%%%%%%%%%%%%%%%%%%%%
\section{Architecture and Implementation Principles}\label{sec:arch}
%%%%%%%%%%%%%%%%%%%%%%%%%%%%%%%%%%%%%%%%%%%%%%%%%%%%%%%%%%%%%%%%%%%%%%

We assume familiarity with the basic concepts of answer-set programming (ASP) (for a thorough introduction to the subject, cf.\ Baral~\cite{baral03}).
In brief, an answer-set program consists of rules of the form
\[
a_1 \vee \cdots \vee a_l \ruleimp a_{l+1}, \ldots, a_{m}, \naf\ a_{m+1}, \ldots, \naf\ a_{n} , 
\]
where $n \geq m \geq l \geq 0$, ``$\naf$'' denotes \emph{default negation}, and  all $a_i$ are first-order literals (\iec atoms possibly preceded by the \emph{strong negation} symbol $\neg$).
For a rule $r$ as above, the expression left to the symbol ``$\ruleimp\!$'' is the \emph{head}  of $r$ and the expression to the right of ``$\ruleimp\!$'' is the \emph{body} of $r$.
If $n=l=1$,
$r$  is a \emph{fact}; if $r$ contains no disjunction, $r$ is \emph{normal}; and
if $l=0$ and $n>0$,
$r$ is a \emph{constraint}. For facts, the symbol ``$\ruleimp\!$'' is usually omitted.
The \emph{grounding} of a  program $P$ relative to its Herbrand universe
is defined as usual.
An \emph{interpretation} $I$ is a finite and consistent set of ground literals, where consistency means that $\{a,\neg a\}\not\subseteq I$, for any atom $a$. $I$ is an \emph{answer set} of a program $P$ if it is a minimal model of the grounding of the \emph{reduct} of $P$ relative to $I$ (see Baral~\cite{baral03} for details).

\medskip
A key aspect in the design of \sealion is extensibility.
That is, on the one hand, we want to have enough flexibility to handle further ASP languages
such that previous features can deal with them with no or little adaption. On the other hand, we want to provide a powerful API framework that can be used by future features.
To this end, we defined a hierarchy of classes and interfaces that represent
\emph{program elements}, \iec fragments of ASP languages.
This is done in a way such that we can use 
common interfaces and base classes for representing similar program elements of different ASP languages. 
For instance, we have different classes for representing
literals of the \gringo language and literals of the \dlv language in order to be able to handle subtle differences.
For example, in \gringo, a literal can have several other literals as conditions,
\egc
\verb|redEdge(X,Y):edge(X,Y):red(X):red(Y)|.
Intuitively, during grounding, this literal is replaced by
the list of all literals \verb|redEdge(n1,n2)|, where 
\verb|edge(n1,n2)|, \verb|red(n1)|, and \verb|red(n2)| can be
derived during grounding.
As \dlv is unaware of conditions, an object of class \verb|DLVStandardLiteral| has no support for them, whereas a \verb|GringoStandardLiteral| object keeps a list of
condition literals.
Substantial differences in other language features,
like aggregates, optimisation, and filtering support,
are also reflected by different classes for \gringo and \dlv, respectively.
However, whenever possible, these classes are derived from a common base class or share common interfaces.
Therefore, plugins can, for example, use a general interface for aggregate literals to refer to aggregates of both languages.
Hence, current and future feature implementations can make use of
high-level interfaces and stay independent of the concrete ASP language to a large extent.

Also, within the \sealion implementation, the aim is to have independent
modules for different features, in form of Eclipse plugins,
that ensure a well-structured code.
Currently, there are the following plugins: 
(i)~the main plugin, (ii)~a plugin that adapts the ANTLR parsing framework~\cite{antlr} to our needs, (iii)~two solver plugins, one for \gringo/\clasp and one for \dlv,
and (iv)~the \kara plugin for answer-set visualisation and visual editing.
Moreover, it is a key aim to smoothly integrate \sealion in the Eclipse platform and to
make use of functionality the latter provides wherever suitable.
The motivation is to exploit the rich platform as well as to ensure compatibility with
upcoming versions of Eclipse.

The decision to build on Eclipse,
rather than writing a stand-alone application from scratch, has many benefits.
For one, we profit from software reuse as we 
can make use of the general GUI of Eclipse and
just have to adapt existing functionality to our needs.
Examples include the text editor framework, 
source-code annotations, problem reporting and quick fixes,
project management, the undo-redo mechanism, the console view,
the navigation framework (Outline, Project Explorer), and
launch configurations.
Moreover, much functionality of Eclipse can be used without any adaptions, e.g.,
workspace management, the possibility to define working sets, \iec grouping arbitrary files and resources together, software versioning and revision control (e.g., based on SVN or CVS),
and task management.
Another clear benefit is the popularity of Eclipse among software developers,
as it is a widely used standard tool for developing Java applications.
Arguably, people who are familiar with Eclipse and basic ASP skills
will easily adapt to \sealion.
Finally, choosing Eclipse for an IDE for ASP offers a chance for integration of development tools for hybrid languages, \iec combinations of ASP and procedural languages.
For instance, \gringo supports the use of functions written in the LUA scripting language~\cite{lua2006}.
As there is a LUA plugin for Eclipse available,
one can at least use that in parallel with \sealion,
however there is also potential for a tighter integration of the two plugins.

The sources of \sealion are available for download from
\begin{quote}
\url{http://sourceforge.net/projects/mmdasp/}.
\end{quote}
An Eclipse update site will be made available as soon as \sealion reaches beta status.

%%%%%%%%%%%%%%%%%%%%%%%%%%%%%%%%%%%%%%%%%%%%%%%%%%%%%%%%%%%%%%%
\section{Current Features}\label{sec:current}
%%%%%%%%%%%%%%%%%%%%%%%%%%%%%%%%%%%%%%%%%%%%%%%%%%%%%%%%%%%%%%%

\begin{figure}[t!]\centering
\includegraphics[width=14.35cm]
% [scale=0.55]
{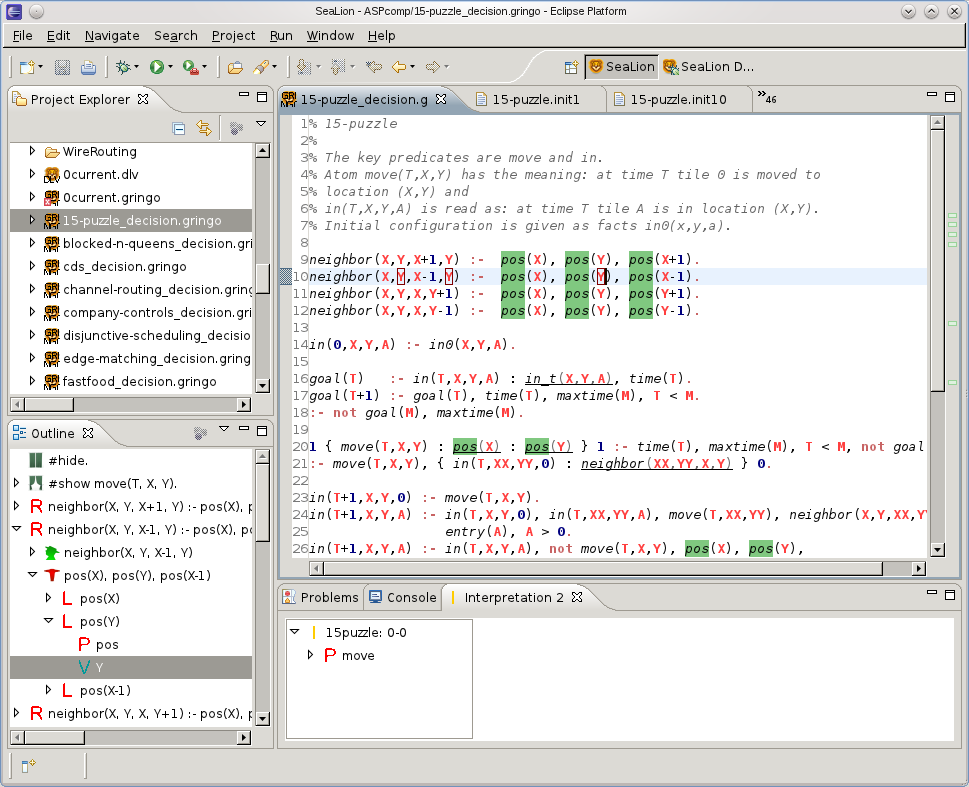}
\caption{A screenshot of \sealion's editor, the program outline, and the interpretation view.}
\label{fig:screenshot}
% \vspace{-3em}
\end{figure}

In this section, we describe the features that are
already operational in \sealion, including technical details on the implementation.

\subsection{Source-Code Editor}

The central element in \sealion is the \emph{source-code editor} for logic programs.
For now, it comes in two variations, one for \dlv and one for \gringo.
A screenshot of a \gringo source file in \sealion's editor is given in Fig.~\ref{fig:screenshot}.
By default, files with names ending in ``.lp'', ``.lparse'', ``.gr'', or ``.gringo''
are opened in the \gringo editor, whereas files with extensions ``.dlv'' or ``.dl''
are opened in the \dlv editor.
Nevertheless, any file can be opened in either editor if required.

The editors provide \emph{syntax highlighting}, which is computed in two phases.
Initially, a fast syntactic check provides initial colouring and styling for
comments and common tokens like dots concluding rules and the rule implication symbol.
While editing the source code, after a few moments of user inactivity,
the source code is parsed and data structures representing the program are computed and stored
for various purposes.
The second phase of syntax highlighting is already based on this program representation
and allows for fine-grained highlighting depending not only on the type of the program element
but also on its role. For instance, a literal that is used in the condition of another literal is highlighted in a different way than stand-alone literals.

The parsers used are based on the ANTLR framework~\cite{antlr} and are in some respect \emph{more lenient
than the respective solver parsers}.
For one thing, they are more tolerant towards syntax errors.
For instance, in many cases they accept terms of various types (constants, variables, aggregate terms) where a solver requires a particular type, like a variable.
The errors will still be noticed, during building the program representation
or afterwards, by means of explicit checks.
This tolerance allows for more specific warning and error reporting than provided by the solvers. For example, 
the system can warn the user that he or she used a constant on the left-hand side of an assignment
where only a variable is allowed.
Another parsing difference is the \emph{handling of comments}.
The parser does not throw them away but collects them and associates them
to the program elements in their immediate neighbourhood.
One benefit is that the information contained in comments can be kept when
performing automatic transformations on the program, like rule reorderings or translations to
other logic programming dialects.
Another advantage is that we can make use of comments for enriching the 
language with our own \emph{meta-statements} that do not interfere with the solver when running the file.
We reserved the token ``\%!'' for initiating meta commands and ``\%*!'' and ``*\%''  for the start and end of block meta commands in the \gringo editor, respectively.
Currently, one type of meta command is supported: assigning properties to program elements.
\begin{example}
In the following source code, a meta statement assigns the name ``r1'' to the rule it precedes.
\begin{verbatim}
%! name = r1; 
a(X) :- c(X).
\end{verbatim}
\end{example}
These names are currently used in a side application of \sealion for reifying disjunctive non-ground programs
as used in a previous debugging approach~\cite{Oetsch-etal09-debug}.
Moreover, names assigned to program elements as above can be seen in Eclipse's ``Outline View''.
\sealion uses this view to give an overview of the edited program
in a tree-shaped graphical representation.
The rules of the programs are represented by the nodes of depth 1 of this tree.
By expanding the ancestor nodes of an individual rule,
one can see its  elements, \iec head, body, literals, predicates, terms, etc.
Clicking on such an element selects the corresponding program code in the editor, 
and the programmer can proceed editing there.
A similar outline is also available in Eclipse's ``Project Explorer'',
as subtree under the program's source file.

%%%%%%%%%%%%%%%%%%%%%%%%%%%%%%%%%%%%%%%%%%%%%%%%%%%
\begin{figure}[t!]\centering
\includegraphics[width=14.65cm]{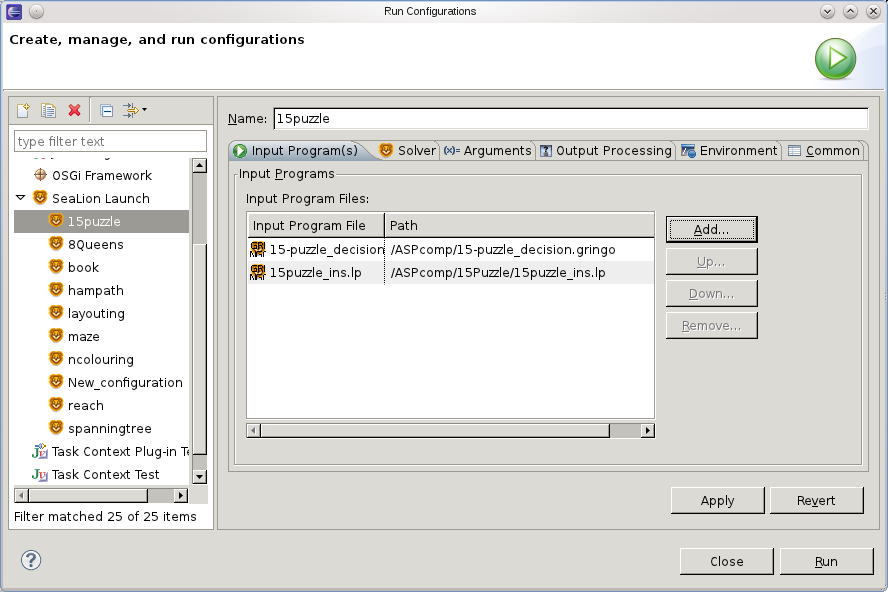}
\caption{Selecting two source files for ASP solving in Eclipse's launch configuration dialog.}
\label{fig:launchconfig}
\end{figure}
%%%%%%%%%%%%%%%%%%%%%%%%%%%%%%%%%%%%%%%%%%%%%%%%%%%

Another feature of the editor is the support for \emph{annotations}.
These are means to temporarily highlight parts of the source code.
For instance, \sealion annotates  occurrences of the program element
under the text cursor. If the cursor is positioned over a literal, all literals of the same
predicate are highlighted in the text as well as in a bar next to the vertical scrollbar
that indicates the positions of all occurrences in the overall document.
Likewise, when a constant or a variable in a rule is on the cursor position,
their occurrences are detected within the whole source code or within the rule, respectively.

Another application of annotations is \emph{problem reporting}.
Syntax errors and warnings are displayed in two ways.
First, as annotations in the source code, they are marked with a zig-zag styled underline.
Second, they are displayed in Eclipse's ``Problem View'' that collects various kinds of problems
and allows for directly jumping to the problematic source code region upon mouse click.

\subsection{Support for External Tools}

In order to interact with solvers and grounders from \sealion,
we implemented a mechanism for handling external tools.
One can define \emph{external tool configurations} that specify the path to an
executable as well as default command-line parameters.
Arbitrary command-line tools are supported; however, there are special
configuration types for some programs such as \gringo, \clasp, and \dlv.
For these, it is planned to have a specialised GUI that allows for a
more convenient modification of command-line parameters.
In addition to external command-line tools, 
one can also define tool configurations that represent pipes between external tools.
This is  needed when grounding and solving are provided by separate executables.
For instance, one can define two separate tool configurations for \gringo and \clasp
and define a piped tool configuration for using the two tools in a pipe.
Pipes of arbitrary length are supported such that arbitrary pre- and post-processing
can be done when needed.

For executing answer-set solvers, we make use of Eclipse's ``launch configuration framework''.
In our setting, a launch configuration defines which programs should be executed using which solver.
Figure~\ref{fig:launchconfig} shows the the page of the launch configuration editor on which input files for a solver invocation can be selected.

%%%%%%%%%%%%%%%%%%%%%%%%%%%%%%%%%%%%%%%%%%%%%%%%%%%
\begin{figure}[t!]\centering
\includegraphics[width=11.65cm]{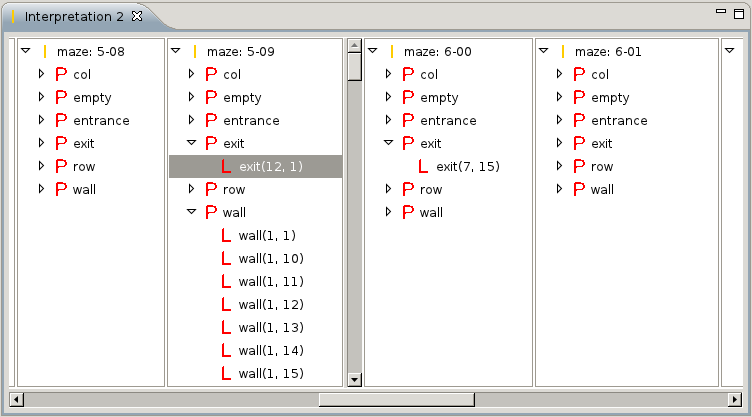}
\caption{\sealion's interpretation view.}
\label{fig:interpretationview}
% \vspace{-1em}
\end{figure}
%%%%%%%%%%%%%%%%%%%%%%%%%%%%%%%%%%%%%%%%%%%%%%%%%%%

Besides using the standard command-line parameters from the tool configurations,
also customised parameters can be set for the individual program launches.

\subsection{Interpretation View}

The programmer can define how the output of an ASP solver run should be treated.
One option is to print the solver output as it is for Eclipse's ``console view''.
The other option is to parse the resulting answer sets and store them in \sealion's \emph{interpretation view}
that is depicted in Fig.~\ref{fig:interpretationview}.
Here, interpretations are visualised as expandable trees of depth 3.
The root node is the interpretation (marked by a yellow ``$I$''), and its children are the predicates (marked by a red ``$P$'')
appearing in the interpretation.
Finally, each of these predicates is the parent node of the literals over the predicate that are contained in the interpretation (marked by a red ``$L$'').
Compared to a standard textual representation, this way of visualising answer sets provides a well-arranged  overview of the individual interpretations.
We find it also more appealing than a tabular representation 
% in the case 
where only entries for a single predicate are visible at once.
Moreover, by horizontally arranging trees for different interpretations next to each other,
it is easy to compare two or more interpretations.

%%%%%%%%%%%%%%%%%%%%%%%%%%%%%%%%%%%%%%%%%%%%%%%%%%%%
\begin{figure}[t!]\centering
\includegraphics[width=14.65cm]{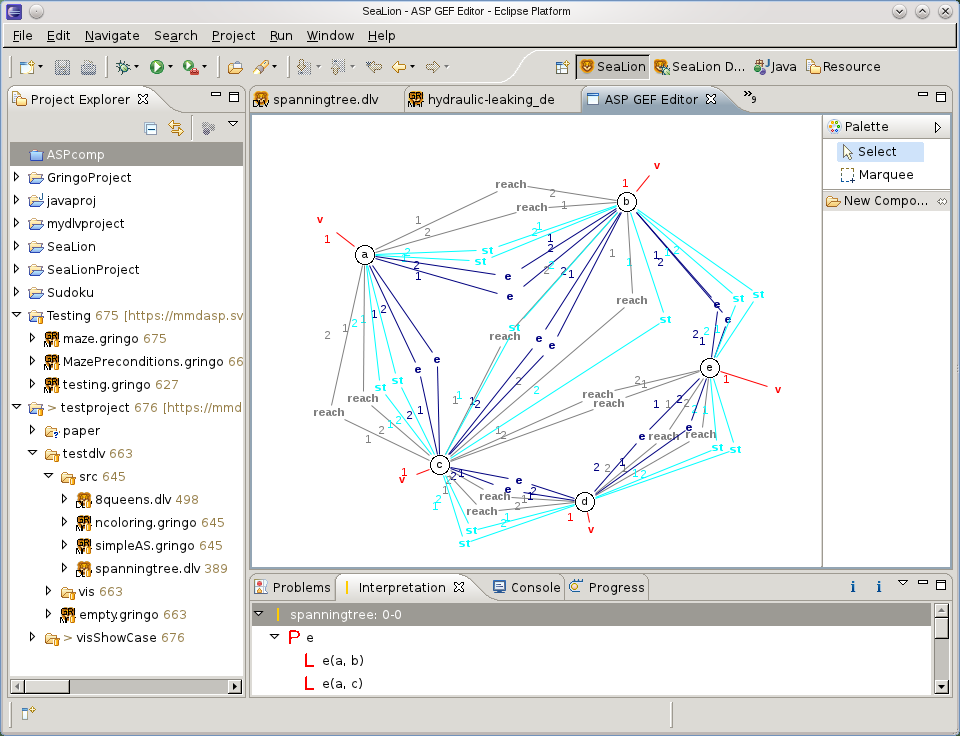}%\label{fig:genericvis}
\caption{A screenshot of \sealion's visual interpretation editor. }
\label{fig:genericvis}
% \vspace{3em}
\end{figure}
%%%%%%%%%%%%%%%%%%%%%%%%%%%%%%%%%%%%%%%%%%%%%%%%%%%%

The interpretation view is not only meant to provide a good visualisation of results,
but also serves as a starting point for ASP developing tools that depend on
interpretations.
One convenient feature is dragging interpretations or individual literals from the interpretation view and dropping them on the source-code editor. 
When released, these are transformed into facts of the respective ASP language.

\subsection{Visualisation and Visual Editing}

The plugin \kara~\cite{kara} is a tool for the graphical visualisation and editing of interpretations.
It is started from the interpretation view.
One can select an interpretation for visualisation by right-clicking it in the view
and choose between a \emph{generic visualisation} or a \emph{customised visualisation}. The latter is
specified by the user by means of a visualisation answer-set program.
The former represents the interpretation as a labelled hypergraph.

In the generic visualisation,
the nodes of the hypergraph are the individuals appearing in the interpretation.
The edges represent the literals in the interpretation, connecting the individuals appearing in the respective literal.
Integer labels on the endings of an edge are used for expressing the argument position of the individual.
In order to distinguish between different predicates, each edge has an additional label stating the predicate name. Moreover, edges of the same predicate are of the same colour. An example of a generic visualisation of a spanning tree interpretation is shown in Fig.~\ref{fig:genericvis}  (the layout of the graph has been manually optimised in the editor).

The customised visualisation feature allows for specifying 
how the interpretation should be illustrated
by means of an answer-set program that uses a powerful pre-defined visualisation vocabulary.
The approach follows the ideas of \aspviz~\cite{CliffeVBP08} and \idpdraw~\cite{idpdraw}: a visualisation program $\Pi_V$ is joined with the interpretation $I$ to be visualised (technically, $I$ is considered as a set of facts) and evaluated using an answer-set solver. 
One of the resulting answer sets, $I_V$, is then interpreted by \sealion for building the graphical representation of $I$.
The vocabulary allows for using and positioning basic graphical elements such as lines, rectangles, polygons, labels, and images, as well as graphs and grids composed of such elements.

The resulting visual representation of an interpretation is shown in a graphical editor that also allows for manipulating the visualisation in many ways.
Properties such as colours, IDs, and labels can be manipulated and
graphical elements can be repositioned, deleted, or even created.
This is useful for two different purposes.
First, for fine-tuning the visualisation before saving it as a scalable vector graphic (SVG) for use outside of \sealion,
using our SVG export functionality.
Second, modifying the visualisation can be used to obtain a modified version $I'$ of the visualised interpretation $I$ by abductive reasoning.

\begin{figure}[t!]\centering
\includegraphics[width=11.65cm]{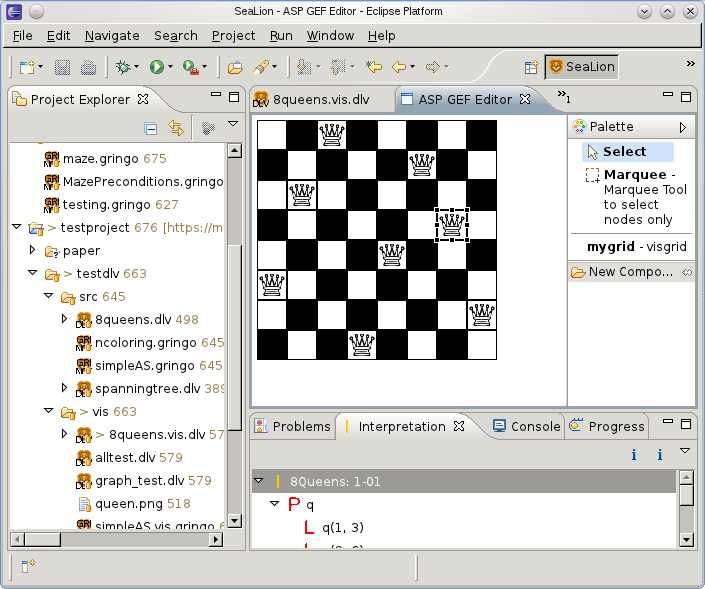}
\caption{A customised visualisation of an 8-queens instance.}
\label{fig:visqueen}
\end{figure}

In fact,  we implemented a feature that allows for abducing an interpretation that would result in the modified visualisation.
Modifications in the visual editor are  automatically reflected in an adapted version $I'_V$
of the answer set $I_V$ representing the visualisation.
We then use an answer-set program $\abdtrans{I'_V}{\Pi_V}$ that is constructed depending
on the modified visualisation answer set $I'_V$ and the visualisation program $\Pi_V$
for obtaining the modified interpretation $I'$ as a projected answer set of $\abdtrans{I'_V}{\Pi_V}$.
For more details, we refer to a companion paper~\cite{kara}.
An example for a customised visualisation  for  a solution to the  8-queens  problem is given in Fig.~\ref{fig:visqueen}.

%%%%%%%%%%%%%%%%%%%%%%%%%%%%%%%%%%%%%%%%%%%%%%%%%%%%%%%%%%%%%%
%%%%%%%%%%%%%%%%%%%%%%%%%%%%%%%%%%%%%%%%%%%%%%%%%%%%%%%%%%%%%%

%%%%%%%%%%%%%%%%%%%%%%%%%%%%%%%%%%%%%%%%%%%%%%%%%%%%%%
\section{Projected Features}\label{sec:future}
%%%%%%%%%%%%%%%%%%%%%%%%%%%%%%%%%%%%%%%%%%%%%%%%%%%%%%

In the following, we give an overview of further functionality that we plan to incorporate into
\sealion in the near future.

One core feature that is already under development
is the support for \emph{stepping-based debugging} of answer-set programs as introduced in recent work~\cite{stepping11}.
Here, we aim for an intuitive and easy-to-handle user interface,
which is clearly a challenge to achieve
for reasons intrinsic to ASP.
In particular, the discrepancy of having non-ground programs
but solutions based on their groundings makes the realisation of practical
debugging tools for ASP non-trivial.

We want to enrich \sealion with support for \emph{typed predicates}.
That is, the user can define the domain for a predicate.
For instance consider a predicate \verb|age/2| stating the age of a person.
Then, with typing, we can express that for every atom \verb|age(t1,t2)|,
the term \verb|t1| represents an element from a set of persons,
whereas \verb|t2| represents an integer value.
Two types of domain specifications will be supported, namely direct ones, which explicitly
state the names of the individuals of the domain, and indirect ones that allow for
specifications in terms of the domain of other predicates.
We expect multiple benefits from having this kind of information available.
First, it is useful as a documentation of the source code.
A programmer can clearly specify the intended meaning of a predicate
and look it up in the type specifications.
Moreover, type violations in the source code of the program can be automatically
detected as illustrated by the following example.
\begin{example}
Assume we want to define a rule deriving atoms with predicate symbol \verb|serves/3|,
where \verb|serves(R,D,P)| expresses that restaurant \verb|R| serves dish \verb|D|
at price \verb|P|. Furthermore, the two predicates \verb|dishAvailable/2| and \verb|price/3|
state which dishes are currently available in which restaurants and the
price of a dish in a restaurant,  respectively.
Assume we have type specifications
stating that for \verb|serves(R,D,P)| and \verb|dishAvailable(D,R)|,
\verb|R| is of type \verb|restaurant| and \verb|D| of type \verb|dish|.
Then, a potential type violation in the rule
\begin{verbatim}
serves(R,D,P) :- dishAvailable(R,D),price(R,D,P)
\end{verbatim}
could be detected, where the programmer mixed up the order 
of variables in \verb|dishAvailable(R,D)|.
\end{example}
In order to avoid problems like in the above example in the first place, autocompletion
functionality could be implemented such that variables and constants of correct types are suggested when writing the arguments of a literal in a rule.
Technically, we plan to realise type definitions within program comments,
similar to other meta-statements as sketched in Section~\ref{sec:current}.

We want to combine the typing system with functionality that allows for defining \emph{program signatures}.
One application of such signatures is for specifying  the predicates and terms used for abducing a modified interpretation $I'$ in our plugin for graphically editing interpretations.
Moreover, input and output signatures can be defined for 
uniform problem encodings, \iec answer-set programs that expect a set of facts representing a problem instance as input such that its answer sets correspond to the solutions for this instance.
Then, such signatures can be used in our planned support for \emph{assertions}
that will allow for defining pre- and post-conditions of answer-set programs.
Having a full specification for the input of a program, \iec a typed signature and
input constraints in the form of preconditions, one can automatically
generate input instances for the program and use them, e.g., for random testing~\cite{testeval11}.
Also, more advanced testing and verification functionality can be realised,
like the automatic generation of valid input (with respect to the pre-conditions) that violates a post-condition.

In order to reduce the amount of time a programmer has to spend for writing type
and signature definitions, we want to explore methods for
partially extracting them from the source code or from interpretations.

Other projected features include typical amenities of Eclipse editors
such as 
refactoring, autocompletion, pretty-printing,
and providing quick-fixes for typical problems in
the source code.
Moreover, checks for errors and warnings that are not already detected by the parser, for example for detecting unsafe variables, need still to be implemented.

We also want to provide different kinds of program translations in \sealion.
To this end, we already implemented a flexible framework for transforming 
program elements to string representations following different strategies.
In particular, we aim at translations between different solver languages 
at the non-ground level.
Here, we first have to investigate strategies when and how transformations of, \egc 
aggregates can be applied such that a corresponding overall semantics can be achieved.
Other specific program translations that we consider for implementation
would be necessary for realising the import and export of rules in the 
Rule Interchange Format (RIF)~\cite{riffld} which is a W3C recommendation for exchanging rules in the context of the Semantic Web.
Notably, a RIF dialect for answer-set programming, called RIF-CASPD, has been proposed~\cite{rifasp}.

Further convenience improvements regarding the use of external tools in \sealion
include the support for setting default solvers for different languages and
a specialised GUI for choosing the command-line parameters.
For launch configurations, we want to add the possibility to directly write the output
of a tool invocation into a file and to allow for exporting the launch configuration as native stand-alone scripts.

Finally, there are many possible ways to enhance the GUI of \sealion.
We want to extend the support for drag-and-drop operations such that, e.g., program elements
in the outline can be dragged into the editor.
Moreover, we plan to realise sorting and filtering features for the outline and interpretation view.
Regarding interpretations, we aim for supporting textual editing of interpretations directly in the
view, besides visual editing, and a feature for comparing multiple interpretations by highlighting their differences.

\section{Related Work}\label{sec:related}
In this section, we give a short overview of existing IDEs for core ASP languages.
To begin with, the tool \ape that has been developed at the University of Bath~\cite{apesea07}
is also based on Eclipse. It supports the language of \lparse and provides
syntax highlighting, syntax checking, program outline, and launch configuration.
Additionally, \ape has a feature to display the predicate dependency graph of a program.
\aspide, a recent IDE for \dlv programs~\cite{aspide11}, is a standalone tool that already offers many features as it 
builds on previous tools~\cite{drawasp10,dlvtracer,spock07}.
Some functionality we want to incorporate in \sealion is already
supported by \aspide, \egc code completion, refactoring, and quick fixes.
Further features of \aspide are support for code templates and a visual program editor.
We do not aim for comprehensive visual source-code editing in \sealion
but consider the use of program templates that allow for expressing common programming patterns.
One disadvantage of \aspide is that the tracing component of the IDE~\cite{dlvtracer} is not publicly available.
In their current releases, neither \ape nor \aspide support graphical visualisation or visual editing of answer sets
as available in \sealion.
\aspide allows for displaying answer sets in a tabular form. 
This is an improvement compared to the standard textual representation but comes with the drawback that only entries for a single predicate are visible at once.
Besides the graphical representation, \sealion can display interpretations
in a dedicated view that gives a good overview of the individual interpretations
and allows also to compare different interpretations.

Concerning supported ASP languages,
\sealion is the first IDE to support the language of \gringo, rather than its \lparse subset.
Moreover, other proposed IDEs for ASP do only consider the language of either \dlv or \lparse,
with the exception of \igrom that
provides basic syntax highlighting and syntax checking for the languages of both, \lparse and \dlv~\cite{igrom}.
Note that \igrom has been developed at our department independently from \sealion
as a student project.
A speciality of \igrom is the support for the front-end languages for planning and diagnosis
of \dlv.
There also exist proprietary IDEs for
ASP related languages with support for object-oriented features,
\OntoStudio and \OntoDLV~\cite{ontostudio2010,ontodlv}.

Compared to \aspviz~\cite{CliffeVBP08} and \idpdraw~\cite{idpdraw},
our plugin \kara~\cite{kara} allows not only for visualisation
of an interpretation but also for visually editing the graphical representation
such that changes are reflected in the visualised interpretation.
Moreover, \kara offers support for generic visualisation,
automatic layout of graph structures, and special support for grids.

\section{Conclusion}\label{sec:conclusion}
In this paper, we presented the current status of \sealion,
an IDE for ASP languages
that is currently under development.
We discussed general principles that we follow in our implementation
and gave an overview of current and planned features.
\sealion is an Eclipse plugin and supports the ASP languages
of \gringo and \dlv.
The most important step in the advancement of the IDE is the integration
of an easy-to-use debugging system.

\end{document}